\documentclass[a4paper]{jpconf}[10pt]
\usepackage{graphicx}
\usepackage{amssymb}
\usepackage{citesort}

\begin{document}
\title{Static density response of one-dimensional soft bosons across the clustering transition}

\author{M Teruzzi$^{1,2}$, D Pini$^2$, S Rossotti$^2$, D E Galli$^2$, G Bertaina$^2$}
\address{$^1$ International School for Advanced Studies (SISSA), Via Bonomea 265, I-34136 Trieste, Italy}
\address{$^2$ Dipartimento di Fisica, Universit\`a degli Studi di Milano, via Celoria 16, I-20133 Milano, Italy}
\ead{gianluca.bertaina@unimi.it}

\begin{abstract}
One-dimensional bosons interacting via a soft-shoulder potential are investigated at zero temperature. The flatness of the potential at short distances introduces a typical length, such that, at relatively high densities and sufficiently strong interactions, clusters are formed, even in the presence of a completely repulsive potential. We evaluate the static density response function of this system across the transition from the liquid to the cluster liquid phases. Such quantity reveals the density modulations induced by a weak periodic external potential, and is maximal at the clustering wavevector. It is known that this response function is proportional to the static structure factor in the classical regime at high temperature, while for this zero-temperature quantum systems, we extract it from the dynamical structure factor evaluated with quantum Monte Carlo methods.
\end{abstract}

\section{Introduction}
Ensembles of particles, interacting with potentials that are flat at short distances, have the possibility of clustering, at sufficiently high particle densities, even if the interaction is completely repulsive. This is a subject of intense study in classical soft-matter physics, where such kind of potentials may describe an effective interaction between polymers \cite{likos_criterion_2001,mladek_formation_2006,Mladek_ComputerAssemblyClusterForming_2008,Lenz_MicroscopicallyResolvedSimulations_2012}. Recently,  a great deal of interest has arisen in the quantum statistical mechanics of particles interacting with finite-range soft potentials, and the related quantum cluster phases. Supersolid behavior, characterized by the coexistence of crystal and superfluid order, has been investigated for bosons \cite{henkel_threedimensional_2010,cinti_defectinduced_2014,saccani_excitation_2012,ancilotto_supersolid_2013,AngeloneSuperglassPhaseInteractionBlockaded2016}, and a breakdown of Fermi liquid theory has been predicted for fermions \cite{Li_EmergenceMetallicQuantum_2016}. In one dimension (1D), cluster Luttinger liquids (CLL) have been proposed on a lattice \cite{mattioli_cluster_2013,dalmonte_cluster_2015} and in the continuum \cite{Rossotti_QuantumCriticalBehavior_2017}. The best candidates, for the experimental realization of such novel quantum phases, seem to be ultracold Rydberg gases \cite{low_experimental_2012}, which are atoms in highly-excited electronic states, and especially ensembles of \textit{dressed} Rydberg atoms, which are superpositions of the ground state and the above mentioned excited states, coupled via a Rabi process. Their effective interaction is a soft-shoulder potential, with a flat repulsive core of size $R_c$, related to the highly excited orbital, and a repulsive van-der-Waals tail \cite{henkel_threedimensional_2010,pupillo_strongly_2010,cinti_defectinduced_2014,balewski_rydberg_2014,macri_ground_2014,plodzien_rydberg_2017}. Experiments are progressively increasing the coherence time of such systems \cite{jau_entangling_2016,zeiher_manybody_2016,zeiher_coherent_2017}. 

In \cite{Rossotti_QuantumCriticalBehavior_2017}, we have characterized the $T=0$ quantum phase transition \cite{sachdev_quantum_2000} from a liquid to a cluster phase in a bosonic system, at a suitable density, by means of quantum Monte Carlo methods. In particular, we have evaluated the dynamical structure factor $S(q,\omega)$, which offers a picture of how the system responds to small dynamical external density perturbations, such as those coming from scattering experiments, providing the dispersion relation of collective modes. In this article, we instead focus on the static response function $\chi(q)$, which provides information on how the system responds to a static periodic potential. This is relevant for trapped quantum gases, where external perturbations can be modulated quite easily. We also briefly describe some of the features of the liquid to cluster-liquid transition.

\section{Methods}
We study a system of $N$ bosons in 1D at linear particle density $n$, governed by the following Hamiltonian in the continuum:
\begin{equation}\label{eq:hamiltonian}
 H = -\frac{\hbar^2}{2m}\sum_i^N \frac{\partial^2}{\partial x_i^2} + \sum_{i<j} \frac{V_0}{r_{ij}^6+R_c^6} \; ,
\end{equation}
where $x_i$ are the particle coordinates, $r_{ij}=|x_i-x_j|$ the distances, $m$ is the mass and $V_0$ and $R_c$ are the strength and the radius of the soft-shoulder potential $V(r)$. A possible realization of this potential in 1D is given by ultracold gases in optical elongated traps. The transverse degrees of freedom would be frozen in their ground state, due to low temperature and strong trapping. The physical hard-core radius $c$ of the core electrons orbitals could be safely ignored, because of integration along the transverse direction, provided the three-dimensional density is small with respect to $1/c^3$. Anyway, the addition of a sufficiently small hard core $c\ll R_c$ to $V(r)$ should not qualitatively change the conclusions of this work, as discussed, in the higher dimensional case, in \cite{Boninsegni_HardCoreRepulsionSupersolid_2016}.

The low-energy physics of 1D quantum fluids is quite generally described by the Luttinger liquid (LL) paradigm \cite{giamarchi_quantum_2003}, characterized by a phononic dispersion law $\varepsilon(q)=\hbar v q$. For Galilean invariant systems, sound velocity $v$ is related to the Luttinger parameter $K_L$ by $v=\hbar n \pi/(m K_L)$ \cite{haldane_effective_1981}. In order to study the excitations at higher energies and momenta, typically numerical methods are required, and we analyze the system using the path integral ground state (PIGS) quantum Monte Carlo method \cite{sarsa_path_2000,rossi_exact_2009}, which projects a trial wavefunction in imaginary time to the ground state. We simulate up to $N=200$ particles in a segment of length $L=N/n$, using periodic boundary conditions (PBC) (see \cite{Rossotti_QuantumCriticalBehavior_2017}). We have thus access to the exact static structure factor $S(q)$ of the system. Moreover, evaluation of the dynamical structure factor $S(q,\omega) = \int dt \frac{e^{i \omega t}}{2\pi N} \langle e^{itH/\hbar}\rho_q e^{-itH/\hbar} \rho_{-q} \rangle$, where $\rho_q$ is the density operator in momentum space, is provided via analytic continuation, through the genetic inversion via falsification of theories algorithm \cite{vitali_initio_2010,bertaina_onedimensional_2016,motta_dynamical_2016,bertaina_statistical_2017}.

\section{Clustering and phase diagram}

It is convenient to introduce the dimensionless density $\rho=nR_c$ and strength $U=V_0/(E_c R_c^6)$, where $E_c=\hbar^2/m R_c^2$.
The Fourier transform $\tilde{V}(q)$ of the potential has the following form:
\begin{equation}
\tilde{V}(q) = E_c\frac{U}{3} \sqrt{\frac{\pi}{2}} e^{-|q|R_c/2} \left[\cos\left(\frac{\sqrt{3}}{2}|q|R_c\right) + e^{-|q|R_c/2} + \sqrt{3} \sin\left(\frac{\sqrt{3}}{2} |q|R_c\right)\right] \; .
\end{equation}
It features a global minimum, found numerically at $q_c \simeq 4.3$, for which $\tilde{V}(q_c)<0$, corresponding to a typical length $b_c=2\pi/q_c \simeq 1.46$ (here and in the following, when not otherwise specified, lengths and wavevectors are in units of $R_c$ and $1/R_c$, respectively).

In the classical case it has been shown that, even for a completely repulsive potential, such a feature favors 
the formation of clusters at a mutual distance $\sim b_{c}$ which, at suitably low temperature $T$ 
or high density $\rho$, arrange into an ordered configuration, thereby forming a cluster crystal 
with periodicity determined by $b_{c}$ itself \cite{likos_criterion_2001,mladek_formation_2006}. To be more specific, in order to have stable clusters, a positive clustering harmonic constant (see \cite{neuhaus_phonon_2011,Rossotti_QuantumCriticalBehavior_2017}) is needed. For such condition to be satisfied it is necessary (but not sufficient) for $\tilde{V}(q)$ to take negative values for some $q$. For the one-dimensional system considered here, this ordered phase is most likely 
prevented by thermal fluctuations at all temperatures 
$T>0$ \cite{prestipino_cluster_2014,prestipino_probing_2015}, but, classically, it still occurs at $T=0$ 
for any positive value of $U$. Indeed, it has been argued that, on increasing $\rho$, one will witness
a virtually infinite sequence of first-order transitions connecting phases with different numbers 
of particles per cluster \cite{neuhaus_phonon_2011,prestipino_probing_2015}.

\begin{figure}[tbp]
\centering
\includegraphics[width = 0.65\columnwidth]{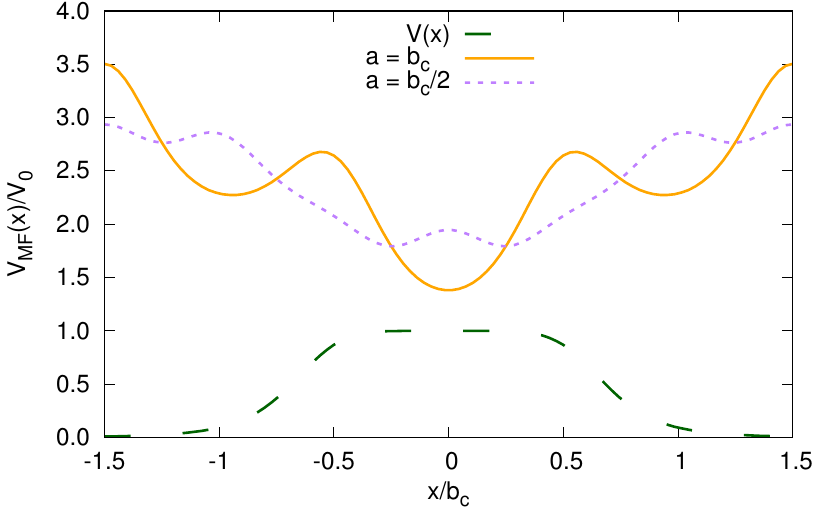}
\caption{Mean-field potentials experienced by a reference particle, at $\rho\simeq 1.37$, if all the others are placed in a lattice of spacing $a=b_c$ (two particles per site) or $a=b_c/2$ (one particle per site). The original soft-shoulder potential is also shown as a reference.}
\label{fig:mf} 
\end{figure}

Let us focus on the density $\rho_2=2/b_c\simeq 1.37$. Classically, at $T=0$, such a density accommodates a crystal of two-particles clusters. This can be easily understood by considering the mean-field potential experienced by a particle, when all the others are placed in couples on a perfect lattice of spacing $a=b_c$, except for a  companion particle, which is placed at $x=0$. The resulting potential (see solid line in Fig.\ref{fig:mf}) manifests a clear harmonic minimum at $x=0$. Notice also that secondary minima are present at multiples of $b_c$, meaning that, upon increasing kinetic energy, hopping to nearby clusters is initially more preferable than melting of the clusters. Conversely, if one forces the other particles to sit on lattice sites with spacing $a=b_c/2$, the mean-field potential displays a double-well structure, close to $x=0$ (dotted line), indicating that such a perfect lattice, with one particle per site, would not be an equilibrium configuration.

\begin{figure}[tbp]
\centering
\includegraphics[width = 0.65\columnwidth]{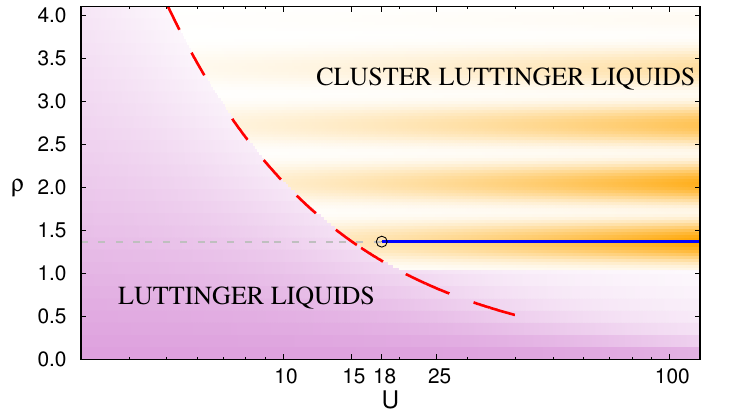}
\caption{Phase diagram (linear-log scale). A symbol indicates the critical point between the LL (dashed line) and CLL (solid line) phases for density $\rho=1.37$ \cite{Rossotti_QuantumCriticalBehavior_2017}, commensurate to two-particle clusters. The long-dashed line corresponds to the softening of the Bogoliubov roton.}
\label{fig:phase} 
\end{figure}

In the quantum case, the scenario is further enriched, since zero-point kinetic energy is present even at $T=0$, introducing two main consequences: the cluster phase is not a crystal anymore, but a liquid of clusters, and clusters can also coherently delocalize towards a standard Luttinger liquid. As a consequence,  
a quantum phase transition occurs between the CLL and a LL without cluster order, at a (density-dependent) value of $U$ (Fig.~\ref{fig:phase}).
This transition has been studied in a related 1D lattice model \cite{mattioli_cluster_2013,dalmonte_cluster_2015}, while we have demonstrated that, at the density $\rho=\rho_2$ in the continuum, it occurs at $U=U_c\simeq 18$ and is in the 2D Ising universality class \cite{Rossotti_QuantumCriticalBehavior_2017}.

In the dilute limit, the only relevant parameter is the scattering length, which we have calculated for the soft-shoulder potential in 1D \cite{teruzzi_microscopic_2017}. At $\rho=\rho_2$, on the contrary, the details of $V(r)$ become relevant. In the LL regime, a Bogoliubov picture turns out to be valid \cite{Rossotti_QuantumCriticalBehavior_2017}, providing the following dispersion relation for the main excited mode
\begin{equation}
\varepsilon_B(q)=\sqrt{\varepsilon_0(q)\left[\varepsilon_0(q)+2 \rho \tilde{V}(q)\right]}  \; ,
\end{equation}
where $\varepsilon_0(q)=\hbar^2q^2/2m$ is the free particle dispersion.
Interestingly, the functional form of the Bogoliubov excitation allows for a minimum of $\varepsilon_B(q)$ at $q\simeq q_c$, provided the combination $\alpha=\rho U$ is sufficiently large. The dynamical structure factors, evaluated in \cite{Rossotti_QuantumCriticalBehavior_2017}, indeed show very clearly that a roton minimum is prominent in the LL regime. When $\alpha=20.65$, the Bogoliubov roton softens, and this mean-field picture breaks down. This can be used as an approximate boundary between the LL and CLL phases (long-dashed line in Fig.~\ref{fig:phase}), although microscopic calculations have to be performed to determine the precise position (see, for example, the circle in Fig.~\ref{fig:phase}, at $\rho=\rho_2$). It is still an open issue the precise determination of the transition between different quantum cluster phases upon increasing $\rho$. 

\section{Static density response function}

\begin{figure}[tbp]
\centering
\includegraphics[width = 0.7\columnwidth]{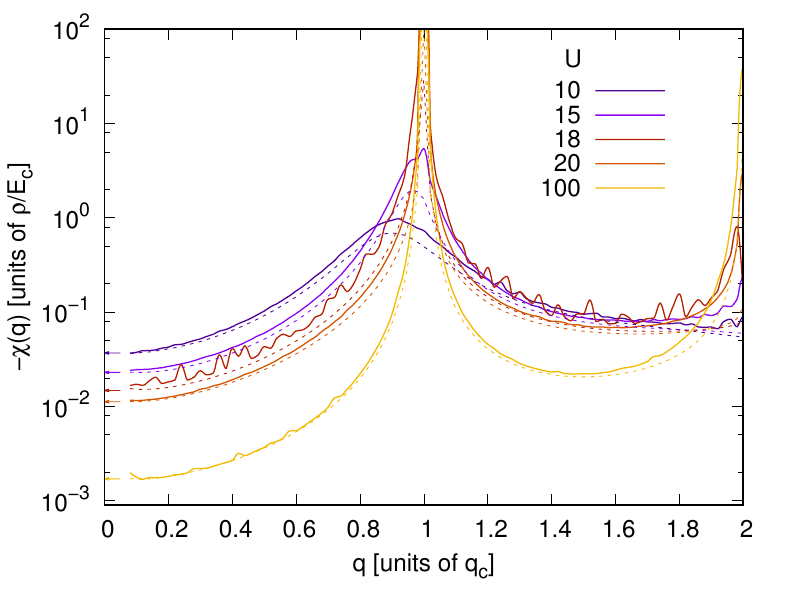}
\caption{Static density response function $\chi(q)$ (with a negative sign), evaluated as (solid lines) the first negative moment of the dynamical structure factor, compared to Feynman single-mode approximation $\chi_{FA}(q)$ (dashed lines), for various $U$ across the transition at $\rho\simeq 1.37$. The arrows indicate the compressibility as extracted from the slope of the static structure factor at at $q\to0$. Errorbars (not shown for clarity) are of the order of the fluctuations of the plotted quantities, and are large only for $U=18$, which approximately corresponds to the transition point and thus suffers more from the finite length of the simulations.}
\label{fig:chiq} 
\end{figure}

The static density response function is introduced, in linear response theory, as the coefficient of proportionality between a weak static periodic perturbation $\tilde{v}(q)$ and the produced density fluctuation $\delta\rho(q)\simeq\chi(q)\tilde{v}(q)$, with respect to the equilibrium homogeneous system \cite{Moroni_StaticResponseLocal_1995,nightingale1998quantum}. At zero temperature, it can be computed by carefully evaluating energy differences between perturbed and unperturbed systems \cite{Moroni_StaticResponseLocal_1995} or from the first negative moment of $S(q,\omega)$ \cite{MottaLinearResponseOneDimensional2017}:
\begin{equation}\label{eq:chiq}
 \chi(q) = - \frac{2\rho}{\hbar} \int_{0}^{\infty} d\omega \, \frac{S(q,\omega)}{\omega} \; .
\end{equation}
The previous equation is the $T\to0$ limit of the more general fluctuation-dissipation relation
\begin{equation}\label{eq:chiqb}
 \chi(q) = - \frac{2\rho}{\hbar} \int_{0}^{\infty} d\omega \, \frac{(1-e^{-\beta\hbar\omega})S(q,\omega)}{\omega} \; ,
\end{equation}
with $\beta=1/k_{\rm B}T$, $k_{\rm B}$ being the Boltzmann constant.
It is straightforward for us to evaluate Eq.~(\ref{eq:chiq}), by integrating $S(q,\omega)$ as obtained in \cite{Rossotti_QuantumCriticalBehavior_2017}. Our results are shown in Fig. \ref{fig:chiq}. We compare $\chi(q)$ from Eq.~(\ref{eq:chiq}) to the Feynman approximation of the same quantity:
\begin{equation}
\chi_{FA}(q)=-2\rho \frac{S(q)^2}{\varepsilon_0(q)} \; .
\end{equation}
The latter assumes that the density fluctuations spectrum is exhausted by a single mode of dispersion $\varepsilon_{FA}(q)$, so that the zeroth
\begin{equation}
 S(q)=\int_0^{\infty}d\omega S(q,\omega)
\end{equation}
and first 
\begin{equation}
 \varepsilon_0(q)/\hbar=\int_0^{\infty}d\omega S(q,\omega)\omega 
\end{equation}
sum rules imply $\varepsilon_{FA}(q)=\varepsilon_0(q)/S(q)$ and $S_{FA}(q,\omega)=S(q)\delta(\omega-\varepsilon_{FA}(q)/\hbar)$, allowing for an estimation of various quantities provided the knowledge of only the static structure factor \cite{Hall_Sumrulesdynamic_1971}. We observe that, qualitatively, $\chi_{FA}(q)$ grasps the main features of $\chi(q)$.

It is worth recalling that, in the classical case, $\chi(q)$ is simply proportional to $S(q)$     
via the relation $\chi(q)=-\beta\rho S(q)$, 
which is recovered from the finite-temperature quantum expression of $\chi(q)$, Eq.~(\ref{eq:chiqb}),
in the classical limit $\beta\hbar\omega\rightarrow 0$.
For the classical fluid, the connection between cluster formation and the occurrence 
of the global minimum of $\tilde{V}(q)$ at $q_{c}\neq 0$ is then brought forward within 
a simple mean-field     
approximation for $S(q)$, according to which $\chi(q)=-\beta\rho/[1+\beta\rho\tilde{V}(q)]$. The minimum 
of $\tilde{V}(q)$ then causes a peak at $q=q_{c}$ in both $S(q)$ and $\chi(q)$, signaling the attitude  
of the system to form density modulations over lengthscales $\sim b_{c}$.   
The above mean-field expression refers to the classical fluid phase and is not valid in the limit
$T\rightarrow 0$. Moreover, in this limit the classical and quantum expressions of $\chi(q)$ 
differ markedly and, as shown by Eq.~(\ref{eq:chiq}), in the latter case the connection between 
$\chi(q)$ and $S(q)$ gets lost. In particular, the quantum $S(q)$ becomes linear in $q$ at small $q$, 
whereas $\chi(q)$ tends to a non-vanishing value which, as in the classical case, is proportional to the
isothermal compressibility of the system $\chi(q){\to} -\rho/mv^2$. \\
Notwithstanding these differences, Fig.~\ref{fig:chiq} clearly shows that, even in the quantum case, 
$\chi(q)$ still displays a peak at $q\sim q_{c}$ which becomes more and more pronounced on increasing $U$,
similarly to what predicted by the above mean-field approximation for the classical $\chi(q)$ 
on increasing $\beta$. 

A comment on finite-size effects is in order. Strictly speaking, $\chi(q)$ (and analogously $S(q)$) may diverge, at a specific wavevector, only in the $N\to \infty$ limit. The number $N=200$ allows us to see a clearly different behavior between $U<U_c$ and $U\ge U_c$: namely, before the transition, the maximum of $\chi(q)$ saturates at a finite value (liquid behavior), while, for stronger interaction, this maximum diverges. It can be shown that such a divergence is sub-linear with the number of particles (quasi-solid behavior, see \cite{Rossotti_QuantumCriticalBehavior_2017} for a detailed analysis of finite-size effects of $S(q)$). Moreover, the number of particles, with PBC, directly sets the momentum resolution $\delta q=2\pi\rho/N$. Finally, since we want to reduce the creation of defects in the two-particles cluster phase, we only simulate an even number of bosons.

The striking feature of this peak of $\chi(q)$ (and $S(q)$) is that its position is determined by the shape of the interaction potential, and not by the mean interparticle distance $1/\rho$, which, for $\rho=\rho_2$, would correspond to $q=2q_c$.  
The static response function we have evaluated thus indicates a very strong sensitivity of the system to external potentials with modulation of wavevector $\sim q_c$, as expected, both in the liquid regime and especially in the cluster regime. By explicitly breaking translational invariance, such perturbations would induce the formation of a true crystal, analog to the superfluid to Mott transition of one-dimensional dilute Bose gases \cite{Stoferle_TransitionStronglyInteracting_2004,Fertig_StronglyInhibitedTransport_2005}. 

We finally discuss the physical understanding of the crucial role of the shape of the potential in favoring clustering. Clearly, the property of the potential of having a negative minimum in its Fourier transform depends on the overall shape of $V(r)$, rather than on its strength $V_0$. A change in $V_0$ would merely amount to a rescaling of the Fourier transform of the interaction, which would affect the depth of its minimum, but would leave its position $q_c$ unchanged.

In principle, the occurrence of a negative minimum in $\tilde V(q)$ may be caused by a number of different features of the interaction. For the potential considered here, it stems from $V(r)$ being rather flat at short distances, while decaying rapidly at longer distances. How such a property may favor cluster formation can be explained heuristically by mean-field arguments similar to that put forward in the discussion of Fig.~\ref{fig:mf}, such as those presented for a classical system in \cite{mladek_multiple_2008}. 
In particular, let us consider a one-dimensional regular array of particles with nearest-neighbor distance $d$, interacting with each other via a repulsive, bounded potential $V(r)$, and let us determine the overall potential $W(r)$ felt by a further test particle inserted into the system for small enough $d$. If $V(r)$ is, say, a Gaussian (whose Fourier transform clearly does not have a negative minimum), then it is found that $W(r)$ has minima in between the particles of the array, suggesting that the preferred position of the additional particle would destroy the periodicity. However, if $V(r)$ is made flatter at small $r$ and steeper at larger $r$, then $W(r)$ has minima on the top of the particles of the array: this would drive the test particle on top of one of these particles, thereby preserving the periodicity by forming a cluster. In such a process, the optimal value of $d$ depends on the detailed shape of $V(r)$, whereas the strength of $V(r)$ does not play any role.  
An extreme (admittedly artificial) example of the latter instance is that of a square-shoulder $V(r)$: when the density of the array reaches the close-packing value such that $d$ is equal to the shoulder width, placing an additional particle in between those of the array will create two overlaps, whereas putting it on top of one of them will create just one. 

We may then say that for this class of potentials, the interaction energy price one has to pay for clustering is more than compensated by the ensuing reduction of the repulsion between particles on neighboring sites. The role of the strength $V_0$ is to tune the relative dominance of potential energy with respect to kinetic energy, and is thus particularly relevant in the 1D quantum case, where it triggers the transition to the liquid phase \cite{Rossotti_QuantumCriticalBehavior_2017}.

\ack
We acknowledge useful discussions with A. Parola. We acknowledge the CINECA awards IscraC-SOFTDYN-2015 and IscraC-CLUDYN-2017 for the availability of high performance computing resources and support. M. T. also acknowledges funding from the European Research Council (ERC) under the European Union's Horizon 2020 research and innovation programme (grant agreement No. 320796-MODPHYSFRICT).

\section*{References}
\bibliographystyle{iopart-num}
\bibliography{proceedings}

\end{document}